\documentclass[prb, twocolumn]{revtex4}

\usepackage[dvips]{color, graphicx}

\usepackage{amsmath, amssymb, amsfonts}

\newcommand{\Msun}{\textrm{M}_{\odot}}

\begin{document}

\title{Hands-on Gravitational Wave Astronomy:\\Extracting
astrophysical information from simulated signals}

\author{Louis J. Rubbo}
\affiliation{Center for Gravitational Wave Physics, Pennsylvania State
  University, University Park, PA 16802}

\author{Shane L. Larson}
\altaffiliation{Formerly at the Center for Gravitational Wave Physics,
  Pennsylvania State University, University Park, PA 16802}
\affiliation{Department of Physics, Weber State University, Ogden, UT
  84408}

\author{Michelle B. Larson}
\altaffiliation{Formerly at the Center for Gravitational Wave Physics,
  Pennsylvania State University, University Park, PA 16802}
\affiliation{Department of Physics, Utah State University, Logan, UT 84322}

\author{Dale R. Ingram}
\affiliation{LIGO Hanford Observatory, Richland, WA 99352}

\date{\today}



\begin{abstract}
  In this paper we introduce a hands-on activity in which introductory
  astronomy students act as gravitational wave astronomers by
  extracting information from simulated gravitational wave signals.
  The process mimics the way true gravitational wave analysis will be
  handled by using plots of a pure gravitational wave signal.  The
  students directly measure the properties of the simulated signal,
  and use these measurements to evaluate standard formulae for
  astrophysical source parameters.  An exercise based on the
  discussion in this paper has been written and made publicly
  available online for use in introductory laboratory courses.
\end{abstract}

\maketitle


\section{Introduction} \label{sec:intro}

Observational astronomy stands at the threshold of an era where
gravitational wave detectors are a tool which regularly contributes
important information to the growing body of astrophysical
knowledge.~\cite{Rubbo:2006}  Ground based detectors such as the Laser
Interferometer Gravitational-wave Observatory~\cite{Abramovici:1992}
(LIGO) and the forthcoming space based detector the Laser
Interferometer Space Antenna~\cite{Sumner:2004} (LISA) will probe
different regimes of the gravitational wave spectrum and observe
sources that radiate at different gravitational wavelengths.  Unlike
their cousins, traditional electromagnetic telescopes, gravitational
wave detectors are not imaging instruments.  How then does a
gravitational wave astronomer take the output from a detector and
extract astrophysical information about the emitting sources?  This
paper introduces a hands-on activity in which introductory astronomy
students answer this question.

Traditional astronomy is often presented through the medium of
colorful images taken with large scale telescopes.  In addition to
studying images, astronomers learn about astrophysical systems by
collecting data at multiple wavelengths, using narrow band spectra,
measuring time varying light curves, and so on.  It is often the case
that the core physics governing the evolution of these distant systems
is deduced from the physical character of the observed electromagnetic
radiation, rather than from the imagery that is used to illustrate the
science for other audiences.

Gravitational wave astronomy is completely analogous to its
electromagnetic cousin, with one important distinction: there will be
no image data, because gravitational wave detectors are not imaging
instruments.  Gravitational wave observatories like LIGO and LISA
return a noisy time series that has encoded within it gravitational
wave signals from one or possibly many overlapping sources.  To gain
information about the systems emitting these gravitational wave
signals requires the use of time series analysis techniques such as
Fourier transforms, Fisher information matrices, and matched
filtering.  Recently an activity has shown how students can emulate
the match filtering process by comparing ideal signals to mocked noisy
detector output.~\cite{Larson:2006}  In this paper, a procedure is
described whereby students can analyze a simulated gravitational wave
signal and extract the astrophysical parameters which describe the
radiating system.  The goal is to introduce students to how
gravitational wave astronomers learn about sources of gravitational
radiation in a fashion suitable for classroom or laboratory exercises
related to this modern and emerging branch of observational
astrophysics.

The rest of this article introduces some basic background of
gravitational radiation and then describes the activity.
Section~\ref{sec:theory} outlines the theory connecting the structure
of gravitational waves to astrophysical parameters, and
Section~\ref{sec:waveforms} illustrates the characteristic waveforms
from a typical binary system.  Section~\ref{sec:procedure} illustrates
a procedure where measurements made from waveform plots, together with
the theory of waveform generation, can be used to extract the
astrophysical parameters (orbital period, distance, etc.) of the
system emitting gravitational radiation.  Section~\ref{sec:discussion}
discusses implementations and extensions for this activity in an
introductory astronomy course.  The analysis described in the paper
has been implemented in an activity format, complete with a keyed
solution for the instructor, and is publicly available
online.~\cite{webpage}

\section{GRAVITATIONAL WAVE PRODUCTION IN BINARIES} \label{sec:theory}

In electromagnetism radiation is produced by an accelerating charged
particle.  Similarly in general relativity, gravitational radiation is
produced by an accelerating mass.  To be precise, gravitational waves
are produced by a time varying mass quadrupole moment.  The reason for
this is straightforward.~\cite{Schutz:1984} Monopole radiation is
prevented due to conservation of mass, while dipole radiation does not
occur due to the conservation of momentum.  This leaves the quadrupole
as the leading order term in the multipole expansion of the radiation
field.  A simple and common example of an astrophysical system with a
time varying quadrupole moment is a binary star system.

For a circular binary system, where the components are treated as
point-like particles, the gravitational waveforms take on the
seductively simple form
\begin{equation} \label{eq:mono}
   h(t) = \mathcal{A}(t) \cos ( \Phi(t) ) \,,
\end{equation}
where $h(t)$ is the gravitational waveform (also referred to as the
gravitational wave strain), $\mathcal{A}(t)$ is the time dependent
amplitude, and $\Phi(t)$ is the gravitational wave phase.  The
amplitude $\mathcal{A}(t)$ can be expressed in terms of the physical
parameters characterizing the system,
\begin{equation} \label{eq:amplitude}
   \mathcal{A}(t) = \frac{2 (G \mathcal{M})^{5/3}}{c^{4}r} \left(
   \frac{\pi}{P_{gw}(t)} \right)^{2/3} \,,
\end{equation}
where $G$ is Newton's gravitational constant, $c$ is the speed of
light, $r$ is the luminosity distance to the binary, and $P_{gw}(t)$
is the gravitational wave period.  The quantity $\mathcal{M} \equiv
(M_{1} M_{2})^{3/5}(M_{1}+M_{2})^{-1/5}$ is called the chirp mass and
appears repeatedly in gravitational wave physics, making it a natural
mass scale to work with.  The origin for this nomenclature will become
evident shortly.  The waveform phase $\Phi(t)$ is given by a the
integral:
\begin{equation} \label{eq:phase}
   \Phi(t) = \Phi_{0} + 2 \pi \int_{0}^{t} \frac{dt'}{P_{gw}(t')} \,,
\end{equation}
where $\Phi_{0}$ is the initial phase.

As in electromagnetism, gravitational waves have two independent
polarization states.  For a binary system the two states are related
by a $90^{\circ}$ phase shift.  Consequently, Eq.~\eqref{eq:mono}
captures the functional form for both polarization states.  For the
purposes of this paper only a single polarization state and its
associated waveform will be discussed.

Gravitational waves carry energy and angular momentum away from the
binary system causing the orbital period to decrease with time
according to~\cite{Hartle:2003}
\begin{equation} \label{eq:orbper}
   P_{orb}(t) = \left( P_{0}^{8/3} - \frac{8}{3} k t \right)^{3/8} \,,
\end{equation}
where $P_{0}$ is the orbital period at time $t = 0$, and $k$ is an
evolution constant given by
\begin{equation} \label{eq:evconst}
   k \equiv \frac{96}{5}  (2 \pi)^{8/3} \left (\frac{G
   \mathcal{M}}{c^{3}} \right)^{5/3} \,.
\end{equation}
As a consequence of the ever shortening orbital period, the two binary
components will slowly inspiral, eventually colliding and coalescing
into a single remnant.  Under the assumption of point-like particles
made here, this formally occurs when $P_{orb}(t) = 0$.

Note that Eq.~\eqref{eq:orbper} gives the orbital period \textit{not}
the gravitational wave period.  Careful scrutiny of
Eqs.~\eqref{eq:amplitude} and~\eqref{eq:phase} will reveal that the
gravitational wave period $P_{gw}(t)$ is the quantity which appears in
the description of the waveform.  Fortunately, for circularized binary
systems, $P_{orb}(t)$ and $P_{gw}(t)$ are simply related:
\begin{equation} \label{eq:periods}
  P_{orb}(t) = 2 P_{gw}(t) \,.
\end{equation}
The simple factor of two stems from the fact that the lowest possible
order for gravitational radiation production is the quadrupole order.
Moreover, quadrupole moments are invariant under a $180^{\circ}$
rotation, yielding a factor of two per complete orbit.

\section{WAVEFORMS FROM A BINARY SYSTEM} \label{sec:waveforms}

As an illustrative example of the kind of waveforms we expect from
binaries, consider a binary neutron star system with $M_{1} = M_{2} =
1.4~\Msun$ ($\mathcal{M} = 1.22 M_{\odot}$) located at the center of
the galaxy $r = 8$~kpc away.  For the activity we will consider
waveforms generated at two distinct times in the binary's evolution.
The first waveform we will consider is $\sim\!10^{6}$ years before
coalescence.  During this phase the gravitational wave frequency is in
the regime that will be detectable by the spaceborne LISA observatory,
which has a principle sensitivity in the range of $10^{-5}$~Hz to
1~Hz.  The second waveform considered will be during the final second
before the neutron star binary coalesces.  The gravitational wave
frequencies during this phase are in the regime that will be
detectable by the terrestrial LIGO observatory, which is sensitive to
gravitational wave frequencies between 10~Hz and $10^{3}$~Hz.

\subsection{Far from Coalescence} \label{sub:far}

\begin{figure}[!tb]
  \begin{center}
    \includegraphics[width=0.45\textwidth]{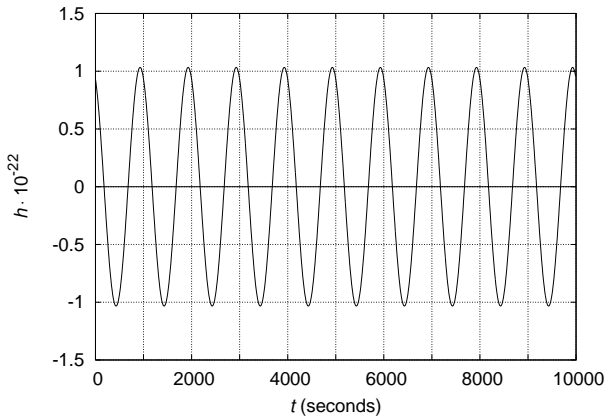}
  \end{center}
  \caption{The gravitational waveform for a binary system consisting of
    two neutron stars far from coalescence and located at the center of
    our galaxy.}
  \label{fig:mono_signal}
\end{figure}

Figure~\ref{fig:mono_signal} shows the emitted gravitational radiation
long before the binary components coalesce.  During this era of the
binary evolution, the gravitational waves are essentially
monochromatic; the orbital period is evolving too slowly to detect a
frequency derivative term.

For monochromatic signals like this, the only measurable properties of
the gravitational waveform are the period $P_{gw}$ (and the orbital
period $P_{orb}$ through Eq.~\eqref{eq:periods}), amplitude
$\mathcal{A}$, and initial phase $\Phi_{0}$.  Even though the waveform
equations depend on the chirp mass $\mathcal{M}$ and the luminosity
distance $r$, it is not possible to solve for their values from the
data provided by the monochromatic waveform.  Not enough information
exists to completely solve Eqs.~\eqref{eq:amplitude} and
\eqref{eq:orbper} together for both quantities.  This can be seen by
considering the relative size of the two terms in
Eq.~\eqref{eq:orbper}; using the binary neutron star chirp mass
$\mathcal{M} = 1.22~\Msun$, it should be evident that the second term
is completely negligible compared to the period $P_{0}$ of the wave
shown in Fig.~\ref{fig:mono_signal}.  In the parlance of gravitational
wave astronomy, there is a \textit{mass-distance degeneracy} in the
waveform description, analogous to the familiar mass-inclination
degeneracy in the electromagnetic observations of spectroscopic
binaries.  This degeneracy is a well known problem, but as the next
section shows, it can be broken if the orbital period of the binary
evolves during the gravitational wave observations.

\subsection{Near Coalescence}\label{sub:near}

\begin{figure}[!tb]
  \begin{center}
    \includegraphics[width=0.45\textwidth]{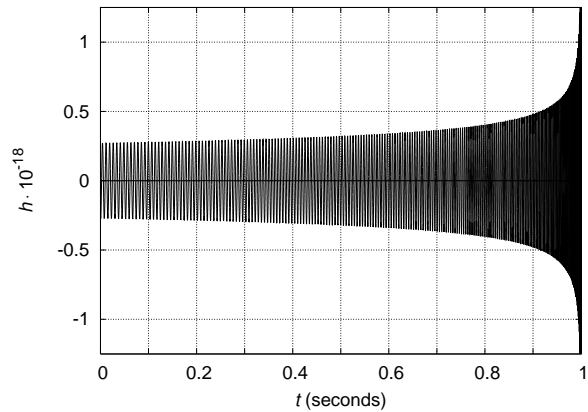}
  \end{center}
  \caption{The waveform over the last second before coalescence.  Since
    the signal's amplitude and frequency is increasing with time, these
    types of systems are said to be \textit{chirping}.}
  \label{fig:full_coal}
\end{figure}

Inspection of Eq.~\eqref{eq:orbper} shows that as time goes on, the
emission of gravitational waves causes the orbital period to grow
shorter, and as a result the frequency of the emitted waves increases.
Similarly, consideration of Eq.~\eqref{eq:amplitude} shows that as the
wave period decreases, the time dependent amplitude $\mathcal{A}(t)$
increases.  This is characteristic behavior for gravitational waves
emitted just prior to a source coalescence, and is known as a
\textit{chirp}. The chirp waveform emitted by the example binary
neutron star system just prior to coalescence is illustrated in
Fig.~\ref{fig:full_coal}.

Any binary signal which evolves appreciably during the gravitational
wave observation is called a chirping binary.  In these cases, the
mass parameter $\mathcal{M}$ which appears in the amplitude
$\mathcal{A}(t)$ and in the period evolution constant $k$ can be
determined from measurements of the evolving signal.  For this reason,
the mass $\mathcal{M}$ is called the \textit{chirp mass}.  To leading
order in gravitational wave production, it is not possible to measure
the individual masses, only the chirp mass.  Consequently, it is not
possible to distinguish between binaries with the same chirp mass.
For example, the binary neutron star considered in this paper with
$M_{1} = M_{2} = 1.4~\Msun$ has roughly the same chirp mass as a
binary with an $M_{1} = 10~\Msun$ black hole and a $M_{2} = 0.3~\Msun$
white dwarf.

To extract the chirp mass from measurements of the gravitational
waveform, consider two small stretches of the chirping waveform.
Figure~\ref{fig:coal_1} shows the waveform from $0~\textrm{s} \leq t
\leq 0.05~\textrm{s}$, and Fig.~\ref{fig:coal_2} shows the waveform
from $0.9~\textrm{s} \leq t \leq 0.92~\textrm{s}$.  The waveform is
appreciably different between these two snapshots, both in amplitude
$\mathcal{A}(t)$ and in period $P_{gw}(t)$.  This allows the
degeneracy found in the monochromatic signal case to be broken,
because the gravitational wave period can be measured at two different
times and used in Eq.~\eqref{eq:orbper} to solve for the chirp mass
$\mathcal{M}$.

\section{MEASURING GRAVITATIONAL WAVEFORMS} \label{sec:procedure}

This section illustrates a procedure at the introductory astronomy
level where students can make direct measurements from the figures in
Section~\ref{sec:waveforms} using a straight edge and the axis labels.
Using their measured data together with the theory presented in
Section~\ref{sec:theory}, the astrophysical character of the system
emitting the gravitational waveforms can be deduced.

\subsection{Monochromatic Waveforms} \label{sub:monochromatic}

Limited astrophysical information can be extracted directly from
Fig.~\ref{fig:mono_signal}, as will be the case with true
monochromatic signals detected by gravitational wave observatories.
With limited assumptions more detailed information can be deduced,
which will be valid so long as the assumptions are valid.  A suitable
extraction and analysis procedure for an introductory astronomy
student would proceed in the following manner:
\begin{itemize}
\item The gravitational wave period $P_{gw}$ can be measured directly
  from the figure.  Since the signal is monochromatic, the binary is
  circular and the orbital period $P_{orb}$ is obtained directly from
  $P_{gw}$ using Eq.~\eqref{eq:periods}.  For the waveform in
  Fig.~\ref{fig:mono_signal} careful measurement should yield a value
  of $P_{gw} = 1000$~sec.
    
\item The amplitude $\mathcal{A}$ and the initial phase $\Phi_{0}$ can
  also be measured directly from the figure.  As noted in
  Section~\ref{sub:far} no astrophysical information can be extracted
  from the amplitude alone.  The initial phase is a simple quantity to
  measure, but does not represent any intrinsic property of the
  binary; its value is solely a consequence of when the gravitational
  wave observations began.  To illustrate this, imagine relabeling the
  time axis in Fig.~\ref{fig:mono_signal} to represent a new
  observation which started somewhat later than the observation shown.
  The initial phase will have some new value, but the waveform itself
  does not change because the intrinsic properties of the binary did
  not change.
    
\item If a gravitational wave astronomer were to assume that the
  binary was a pair of neutron stars, the component mass values could
  be assigned as part of the assumption.  Most neutron star masses
  cluster around $M = 1.4~\Msun$, so a good base assumption is that
  each component of the binary has this mass.  As noted in
  Section~\ref{sub:far} this assumption can be a dangerous one, since
  similar chirp masses $\mathcal{M}$ can result from significantly
  different systems.  Other information, not present in the
  gravitational waveform, may help an astronomer feel more confident
  about such an assumption.  For example, an associated simultaneous
  electromagnetic signal or the location of the source on the sky may
  favor one model of the binary over another.
    
\item If the masses are assumed, the orbital separation of the binary
  components, $R$, can be computed from the measured orbital period by
  using Kepler's Third Law:
  \begin{equation} \label{eq:Kepler3}
    G (m_{1} + m_{2}) = \left( \frac{2\pi}{P_{orb}} \right)^{2} R^{3} \,.
  \end{equation}
  For this example, the orbital separation is $R = 1.4 \times
  10^{-3}~\textrm{AU} = 2.1 \times 10^{8}~\textrm{m}$, or a little
  less than the separation of the Earth and the Moon.

\item If the masses are assumed, the distance to the binary can be
  computed from Eq.~\eqref{eq:amplitude} and the measured amplitude.
  If careful measurements have been made, the answer should be close
  to the value $r = 8~\textrm{kpc} = 2.5 \times 10^{20}~\textrm{m}$.
    
\item Lastly, if the masses are assumed, it can be quantitatively
  shown that the monochromatic descriptor is a good one for this wave
  by computing the value of the second term in Eq.~\eqref{eq:orbper}
  and showing that it is negligible compared to the measured period
  $P_{0}$.
\end{itemize}

\subsection{Chirping Waveforms} \label{sub:chirping}

\begin{figure}[!tb]
  \begin{center}
    \includegraphics[width=0.45\textwidth]{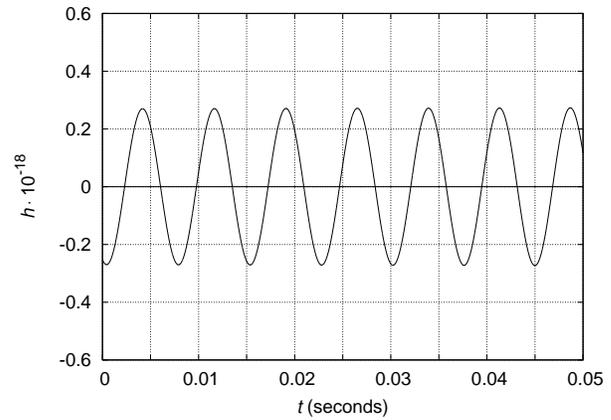}
  \end{center}
  \caption{The chirping waveform one second before coalescence.}
  \label{fig:coal_1}
\end{figure}

\begin{figure}[!tb]
  \begin{center}
    \includegraphics[width=0.45\textwidth]{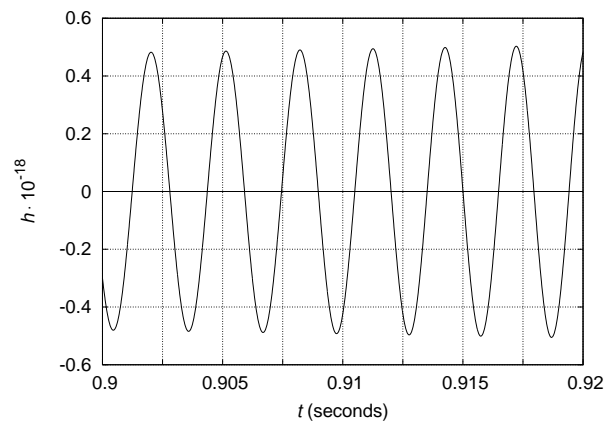}
  \end{center}
  \caption{The chirping waveform one-tenth of a second before
    coalescence.}
  \label{fig:coal_2}
\end{figure}

In the case of a chirping waveform, additional astrophysical
information associated with the system can be extracted directly from
measurements of the waveform without making underlying assumptions
like those needed when the system was far from coalescence.  To
extract information from the chirping waveform shown in
Fig.~\ref{fig:full_coal}, the two zoom-ins of the waveform shown in
Figs.~\ref{fig:coal_1} and \ref{fig:coal_2} will be used.
A typical extraction procedure might look like this:
\begin{itemize}
\item For each of the Figs.~\ref{fig:coal_1} and \ref{fig:coal_2},
  measure the period of one cycle of the wave, and note the time $t$
  at which the periods were measured.  The amplitudes $\mathcal{A}(t)$
  should be measured for the same cycle as the periods.
   
\item If the period measured at time $t_{1}$ in Fig.~\ref{fig:coal_1}
  is $P_{0}$, and the period measured at time $t_{2}$ in
  Fig.~\ref{fig:coal_2} is $P_{gw}(t)$ at time $t = t_{2} - t_{1}$,
  then Eq.~\eqref{eq:orbper} can be used to deduce the chirp mass
  $\mathcal{M}$ of the system.
   
\item Once the chirp mass $\mathcal{M}$ has been determined, the
  distance to the binary can be computed by using
  Eq.~\eqref{eq:amplitude} with the measured amplitude
  $\mathcal{A}(t)$ and period $P_{gw}(t)$ of each waveform.  The
  results from the two figures can be averaged together to obtain a
  final result.
\end{itemize}

\section{Discussion} \label{sec:discussion}

This paper introduced the core calculations in gravitational wave
astrophysics an introductory astronomy student can perform in a
laboratory setting to glean information about an astrophysical system.
To compliment this article we have also developed a student activity
sheet and corresponding teacher's guide related to the exercises
described in sections~\ref{sub:monochromatic} and \ref{sub:chirping}.
The complimentary material is available, along with a template
activity,~\cite{Larson:2006} at
\texttt{http://cgwp.gravity.psu.edu/outreach/activities/}.

The activity described here is a simple introduction to how
gravitational wave astronomers extract astrophysical information from
observed binary waveforms.  Real signal analysis is a more complex
endeavor than what has been presented here.  The most significant
challenge in the case of true data is identifying the signal buried in
a noisy data stream.  A common approach to this problem in
gravitational wave astronomy is to use template matching, which has
been explored in a separate activity.~\cite{Larson:2006} If a signal
is present in a noisy data stream, the template provides a way to
subtract the noise away and leave a clean waveform behind.  This is
the assumed starting point in the activity developed here.  Its from
clean waveforms that astronomers will estimate the values of
astrophysical parameters describing a source of gravitational waves.
By sequencing the two activities a student is exposed, at least in an
idealized way, to the methods used by gravitational wave astronomers
to extract astrophysical information about the emitting systems.


\begin{acknowledgments}
  This work was supported by the Center for Gravitational Wave
  Physics.  The Center for Gravitational Wave Physics is funded by the
  National Science Foundation under cooperative agreement
  PHY-01-14375.  The authors would also like to thank the LIGO
  Laboratory.  LIGO is funded by the National Science Foundation under
  Cooperative Agreement PHY-0107417.
\end{acknowledgments}


\bibliographystyle{apsrev}

\end{document}